\documentclass[aps,prl,twocolumn,superscriptaddress]{revtex4-1}
\usepackage{amsmath}
\usepackage{amsfonts}
\usepackage{amssymb}
\usepackage{graphicx}
\usepackage[english]{babel}
\usepackage[utf8x]{inputenc}
\usepackage[T1]{fontenc}
\usepackage{amsthm}
\usepackage[colorinlistoftodos]{todonotes}
\usepackage[shortlabels]{enumitem}
\usepackage[colorlinks=true, allcolors=blue]{hyperref}
\usepackage{natbib}
\usepackage{tikz}
\usepackage[normalem]{ulem}
\usepackage{comment}

\newcommand{\R}{\mathbb{R}}
\DeclareMathOperator{\sgn}{sgn}    

\begin{document}

\title{Geometric frustration induces the transition between rotation and counterrotation in swirled granular media}

\author{Lisa Lee}
\affiliation{John A. Paulson School of Engineering and Applied Sciences, Harvard University, Cambridge, MA 02138}

\author{John Paul Ryan}
\affiliation{Department of Computer Science, Cornell University, Ithaca, NY 14850}

\author{Yoav Lahini}
\affiliation{Raymond and Beverly Sackler School of Physics and Astronomy, Tel Aviv University}

\author{Miranda Holmes-Cerfon}
\thanks{To whom correspondence should be addressed. E-mail: holmes@cims.nyu.edu, shmuel@seas.harvard.edu}
\affiliation{Courant Institute of Mathematical Sciences, New York University, New York, NY 10012}

\author{Shmuel M. Rubinstein}
\thanks{To whom correspondence should be addressed. E-mail: holmes@cims.nyu.edu, shmuel@seas.harvard.edu}
\affiliation{John A. Paulson School of Engineering and Applied Sciences, Harvard University, Cambridge, MA 02138}

\date{\today}

\begin{abstract}
Granular material in a swirled container exhibits a curious
transition as the number of particles is increased: at low densities the particle
cluster rotates in the same direction as the swirling motion of the container,
while at high densities it rotates in the opposite direction. We
investigate this phenomenon experimentally and numerically using
a co-rotating reference frame in which the system reaches a
statistical steady-state.
In this steady-state the particles form a cluster whose translational degrees of freedom are stationary, while the individual particles constantly circulate around the cluster's center of mass, similar to a ball rolling along the wall within a rotating drum. We show that the transition to counterrotation is friction-dependent.
At high particle densities, frictional effects result in geometric frustration
which prevents particles from cooperatively rolling and spinning.
Consequently, the particle cluster rolls like a rigid body with no-slip conditions on the container wall, which necessarily counterrotates around its own axis.
Numerical simulations verify that both wall-disc friction and
disc-disc friction are critical for inducing counterrotation.
\end{abstract}

\maketitle

From hurricanes to bacterial swarms, the emergence of system-scale circulation from local interactions and local driving is a phenomenon exhibited on many scales and in many different physical systems. In 2D turbulence, vorticity at the small injection scale may cascade to larger and larger scales, stabilizing into a single system-scale vortex \cite{shats2005,xia}. Analogous behavior is observed in an active fluid of spinners flowing though a lattice of annular channels and driven by a magnetic field, which breaks time reversal symmetry. This active liquid develops sound modes that propagate along the boundary, generating global circulation \cite{souslov2017}. Finally, dense suspensions of self-propelled bacteria in confinement also lead to the spontaneous formation of stable circulation along the container walls \cite{wioland2016,lushi2014}. In all of these systems, rotation is actively or passively injected locally, and interactions between the local units ultimately lead to global circulation of the entire system.

There is no need to go to very complex systems to observe the nontrivial emergence of system-size circulation. A handful of marbles swirled in a teacup exhibits similar dynamics. When there are only a few marbles in the container, they form a line that rolls along the container wall at the frequency of the circular translations ("snake" mode \cite{r5}). When a couple more marbles are added, they form a cluster that sloshes periodically around the container, its individual marbles repeatedly cramming against the wall at the outer edges of the swirl before flowing freely through the rest of the container. This sloshing marble ensemble rotates about its own center of mass in the same direction as the container, much like wine swirling in a glass. When even more marbles are added, the monolayer of marbles will continue to periodically slosh around the container, but, curiously, above a critical marble density the pack reverses to counterrotation, or to rotating in the direction \emph{opposite} to that of the container \cite{r2}. 
This transition to counterrotation is startling, because the angular momentum of the marbles in their own reference frame changes sign with the addition of a couple marbles, and is ultimately opposite that of the hand which forces them.
The transition 
is not restricted to marbles in a teacup, but can be observed in a variety of shaken containers filled with granular material, from vibration mills for grinding industrial materials \cite{Yokoyama:1996io} to baby rattles.

Such a simple and ubiquitous phenomenon demands an explanation, yet, while it has been studied empirically \cite{r1,r2,r3,r4,r5}; there is still no clear understanding of the minimal ingredients necessary for it to occur.
We propose to understand this phenomenon via analogies to a single rigid body and a fluid in a swirled container.
A single rigid body in a swirled container with no-slip boundary conditions will roll commensurately on the wall (akin to one gear inside another), rotating about its own center of mass in the opposite direction as the container. One sees similar behavior for a pancake rolling on the edge of a swirled frying pan. In contrast, a rigid body with perfect-slip boundary conditions (as if the pancake were in a bath of melted butter) would receive no torque from the boundary, and hence experience no overall sense of rotation.
Finally, a fluid will rotate in the same direction as the container is swirled, 
 also to varying degrees based on its slip conditions with the container wall.
Thus the rotation behavior of a single body in a swirled container depends on its own rigidity as well as its slip interactions with the wall.
For granular systems composed of many interacting bodies, the fluidity and boundary interactions are not material constants but are emergent dynamic properties \cite{0295-5075-16-3-006,1402-4896-1989-T29-048,RevModPhys.68.1259,Bi2011,majmudar2007}, determined by the friction between the particles as well as the friction with the container wall.

Here we experimentally and numerically investigate the dynamics of swirling particles with the goal of identifying the minimal physical factors that facilitate the transition from rotation to counterrotation as the number of particles is increased.
We show that the critical control parameter for counterrotation is the particle cluster's effective slip with the wall, which arises from friction. Although the individual particles' friction coefficient with the wall does not depend on the number of particles, the effective boundary condition of the particle collective does and can change from perfect slip to perfect stick as the number of particles is increased.
Our experimental observations suggest that friction between the particles prevents densely packed particles from cooperatively rolling and spinning, causing the particle cluster to rigidify and roll on the wall as a whole, ultimately resulting in counterrotation. Finally, we numerically simulate the swirling system and find that if either interparticle friction or particle-wall friction is set to zero, the system rotates at the same frequency at all densities and never transitions to counterrotation, verifying that friction is critical for counterrotation.

\begin{figure}[h]
\includegraphics[width=1\linewidth]{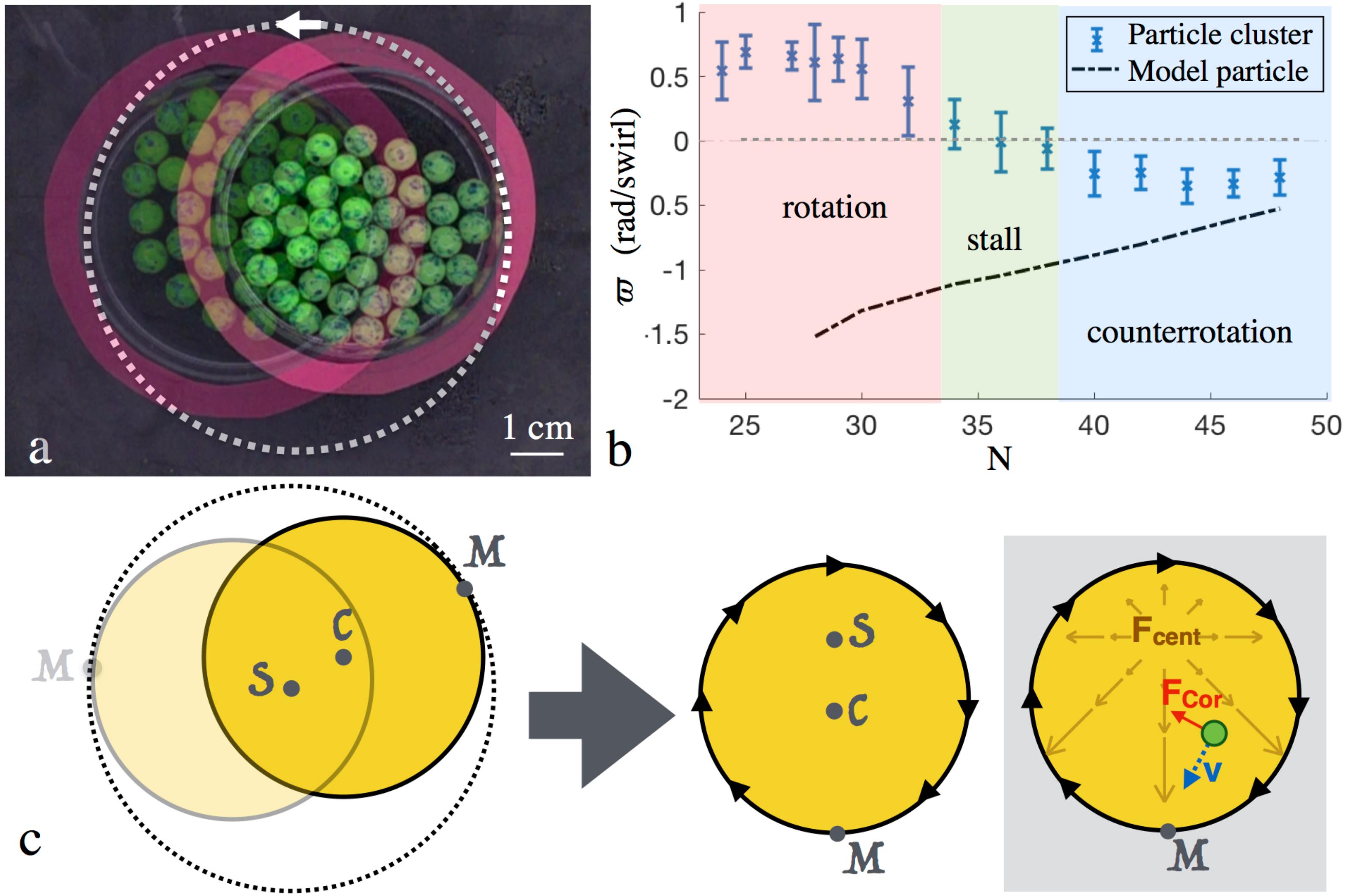}
\caption{
a: Raw image of particles in a swirling container, and the same container at a later time after it has translated through roughly half its circular trajectory. The white dotted line represents the outermost points of the container during its trajectory. The container does not rotate in the lab frame. b: The mean angular velocity ($\varpi$) of the particle cluster about its own center of mass for different particle counts $N$. As  $N$ increases the cluster transitions from rotation to stalling to counterrotation. Black dashed line indicates the angular velocity of a theoretical single particle or pancake, with perimeter matching that of the cluster, if it were perfectly rolling along the wall. c: The M-frame rotates with the container such that the point (left, labeled "M") on the container's boundary furthest from the center of swirling is always positioned at the bottom in the M-frame (middle and right). As a result, the walls of the container in the M-frame appear to be rotating clockwise, as depicted by the black arrows. Also shown are the point $S$ center of swirling, and the point $C$ center of the container. The rightmost M-frame diagram depicts the centrifugal force field (F$_{\text{cent}}$, brown) as well as the Coriolis force (F$_{\text{Cor}}$, red) associated with a particle (green) with indicated velocity ($v$, blue). \label{fig1}}
\end{figure}

Experimentally, the transition from rotation to counterrotation of swirling particles is observed upon a RotoMix orbital table performing circular translations, without rotation, of angular velocity $\omega = 11.81$ rad/sec and amplitude 1.15 cm. The mounted circular container, 5.1 cm in diameter, with a vertical edge is partially filled with a monolayer of $N$ (ranging from 24 to 48) plastic spheres (mass $m=.12$ g and diameter 6 mm) and imaged from above using a stationary Sony RX100 IV camera at 960 fps, as shown for two typical images in Fig \ref{fig1}a. 
We call the collection of particles a ``cluster,'' and measure its average angular velocity $\varpi$ by the
average angular velocity of the particles about the cluster's center of mass: $\varpi= \frac{1}{T-1} \frac{1}{N} \sum_{t=2}^{T}\sum_{i=1}^{N}
\varpi_i(t)$. Here $T$ is the total number of frames and $\varpi_i(t)$ is particle $i$'s instantaneous angular velocity about the cluster's center of mass at time $t$:
$\varpi_i(t)=\frac{1}{\Delta t}(\arctan(\frac{y(t,i)-y_{\text{CM}}}{x(t,i)-x_{\text{CM}}})  - \arctan(\frac{y(t-1,i)-y_{\text{CM}}}{x(t-1,i)-x_{\text{CM}}}) ) $. 
Here $x_{\text{CM}} $ and $y_{\text{CM}} $ refer to the x- and y-coordinates of the center of mass of all particles.

Under these conditions, the average angular velocity $\varpi$  crosses zero and transitions between rotation and counterrotation at roughly 36 particles, as shown in Fig \ref{fig1}b and the SI video. For $N>36$, the cluster counterrotates ($\varpi < 0$), and for even higher values of $N$, the average angular velocity reaches a minimum and then increases slightly for the largest values of $N$.

Further analyzing the data in the lab frame is a challenge since 
the particles experience a rotating time-dependent force from the circular translations of the swirling container; thus the particles slosh around the container with no apparent steady-state (SI Video). Therefore, it is illuminating to analyze the data in a frame of reference where the external forces are steady \cite{r3}. Consider a frame which rotates at the angular velocity $\omega$ of the container, about an axis at the center $S$ of the swirling orbit.
 In this frame of reference the container's translational velocity is zero and it rotates around its own center $C$ at a constant angular speed $-\omega$.  
 This frame of reference is equivalent to centering our camera above the center of the container and rotating it such that $S$ is stationary, keeping the boundary point furthest from $S$ at the bottom; Hereafter we refer to this point as the point M, as shown in Fig \ref{fig1}C. In this frame of reference there are two external forces acting on all particles: the centrifugal force, which points radially away from $S$ with magnitude at a point $x$ equal to $m \omega^2 |x-S|$, and the Coriolis force, which points perpendicularly to the right of the particle's velocity vector $v$, with magnitude $2 m |v| \omega$. Within the container, the centrifugal force pushes the particles outward to the boundary, and at most of the boundary the particles are then pushed towards point M. 
At the boundary, the particles are also dragged clockwise due to friction with the moving container wall. 
We call the original frame of reference the lab frame, and the rotating frame of reference the M-frame. The latter terminology is borrowed from Kumar et al \cite{r3}, which used this frame of reference to study granular particles on the edge of a swirling cylinder.

\begin{figure}[h]
\includegraphics[width=1\linewidth]{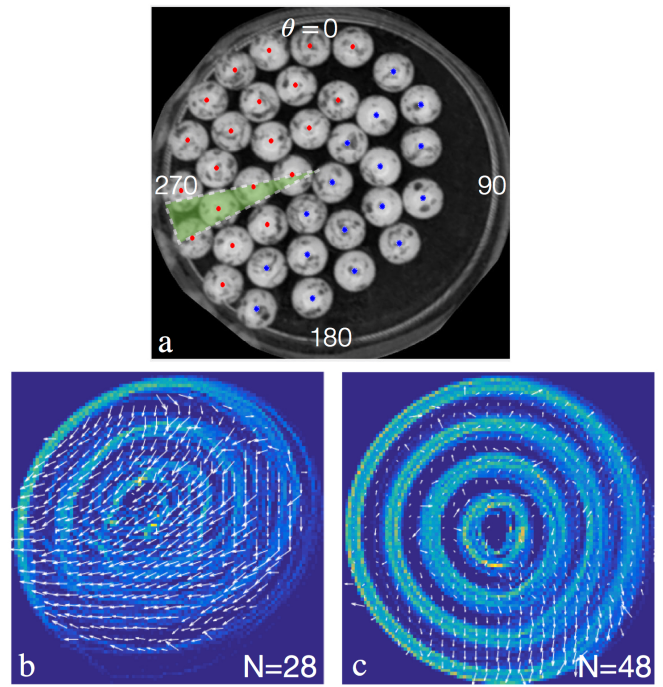}
\caption{
a: Sample image in the M-frame, with pinned (red) and loose (blue) region particle centers labeled. A sample angular slice interrogation area is highlighted in green. The angular slice slides around the entire container during analysis; the major angular positions $\theta$ are denoted on the edge of the container. b: Density histogram of the particles for a rotation case at low $N=28$. The white arrows denote the average local particle velocity deviation from the velocity of the underlying container.  c:  Same as b, but for counterrotation at high $N=48$. The coherent pinned region (top left on the container) grows in size with increasing $N$, while the loose region shrinks in size. The particles perform clockwise trajectories around the container while passing between the pinned and loose regions in a cyclic manner. \label{fig2}}
\end{figure}

In the M-frame, our system appears to be at steady state (SI Videos and Fig \ref{fig2}a), characterized by a steady particle density pattern and  a steady circulation pattern when averaged over typical fluctuations, as shown in Figures \ref{fig2}b and \ref{fig2}c. 
Visually, the dynamics resemble particles in a rotating drum, where the wall rotates at a steady velocity and the beads are uniformly pulled downwards. In the rotating drum, gravity aggregates the beads at the bottom, where the wall then drags the bead cluster upwards until it liquefies and the beads avalanche back to the bottom of the drum, continuing the cycle. In the M-frame, the force of gravity is replaced by the centrifugal force, both tending to push particles toward point M at the bottom. Additionally, the M-frame introduces the Coriolis force that the rotating drum lacks, resulting in the beads tending to the left of the container. 

In the M-frame, all particles follow clockwise trajectories within the container while passing between two distinct spatial regions: a dense, ``pinned'' region (solid-like), and a sparse, ``loose'' region (liquid-like), as shown in Fig \ref{fig2}a. In the pinned region, the particles are packed in layers against the moving wall and move as a rigid structure, whereas in the loose region the particles perform less constrained trajectories as they cross the container, as shown in Fig \ref{fig2}b. When a particle in the loose region transitions to the pinned region, it undergoes multiple collisions before settling into a trajectory parallel to the container's edge. Increasing $N$ also increases the number of collisions a particle experiences, causing it to enter the pinned region earlier in its cycle and subsequently increasing the size of the pinned region. For all values of $N$, particles exit the pinned region and detach from the rigid structure at the same location. Once loose, a particle experiences centrifugal (0-460 cm/s\textsuperscript{2}) and Coriolis (0-200 cm/s\textsuperscript{2}) forces and accelerates as it traverses the loose region. Multiple collisions at the end of the loose region pin the particle and this cycle continues.

Individual particles circulate between these two regions and on average the particle cluster rotates in the same clockwise direction as the container in the M-frame, shown for two typical examples in Fig \ref{fig2}b and \ref{fig2}c. Importantly, the cluster's angular velocity in the M-frame, $\varpi_M$, is related to its angular velocity in the lab frame as 
\begin{equation}\label{labtoM}
\varpi_M = \varpi-\omega\,.
\end{equation} 
Therefore, counterrotation in the lab frame ($\varpi<0$) corresponds to the particle cluster rotating faster than the dish in the M-frame ($|\varpi_M| > \omega$).

\begin{figure}[h]
\includegraphics[width=1\linewidth]{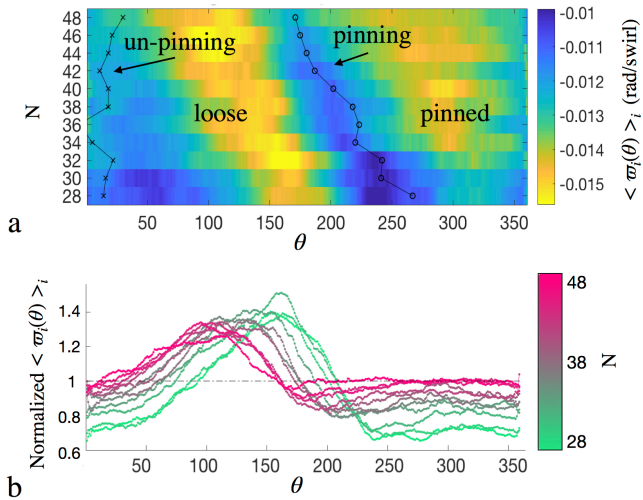}
\caption{a: Average particle angular velocity $\langle\varpi^M_i(\theta)\rangle_i$ about the cluster's center of mass, as a function of $\theta$ and $N$. 
In the loose region, $\varpi_i(\theta)$ increases and decreases in a quantitatively similar manner for all values of $N$, with the only difference being the location on the dish at which this peak occurs. However, in the pinned region, the average value of |$\varpi_i(\theta)$| consistently increases with increasing $N$.  Plotted in black are the average locations at which particles enter the pinned and loose regions. Note that the apparent decrease of $|\langle\varpi^M_i(\theta)\rangle_i|$ in the transition regions is an artifact of the ooblong shape of the particle cluster. The transition regions occur at higher distances from the center, but the linear velocity of the particles during the transition regions does not change much. 
b: Average angular velocity of particles circling the dish, normalized with respect to the angular velocity of the wall about the cluster's center of mass. This is done by dividing $\langle\varpi_i(\theta)\rangle_i$ by the angular velocity of the boundary around the cluster's center of mass, a function of $\theta$ since the center of mass is not at $C$.
At low $N$, a normalized velocity <1 in the pinned region indicates that those particles are traveling slower than the dish wall. As $N$ is increased, the normalized velocity of particles in the pinned region approaches 1, meaning the particles are moving at the velocity of the wall.
\label{fig3}}
\end{figure}

\begin{figure}[h]
\includegraphics[width=1\linewidth]{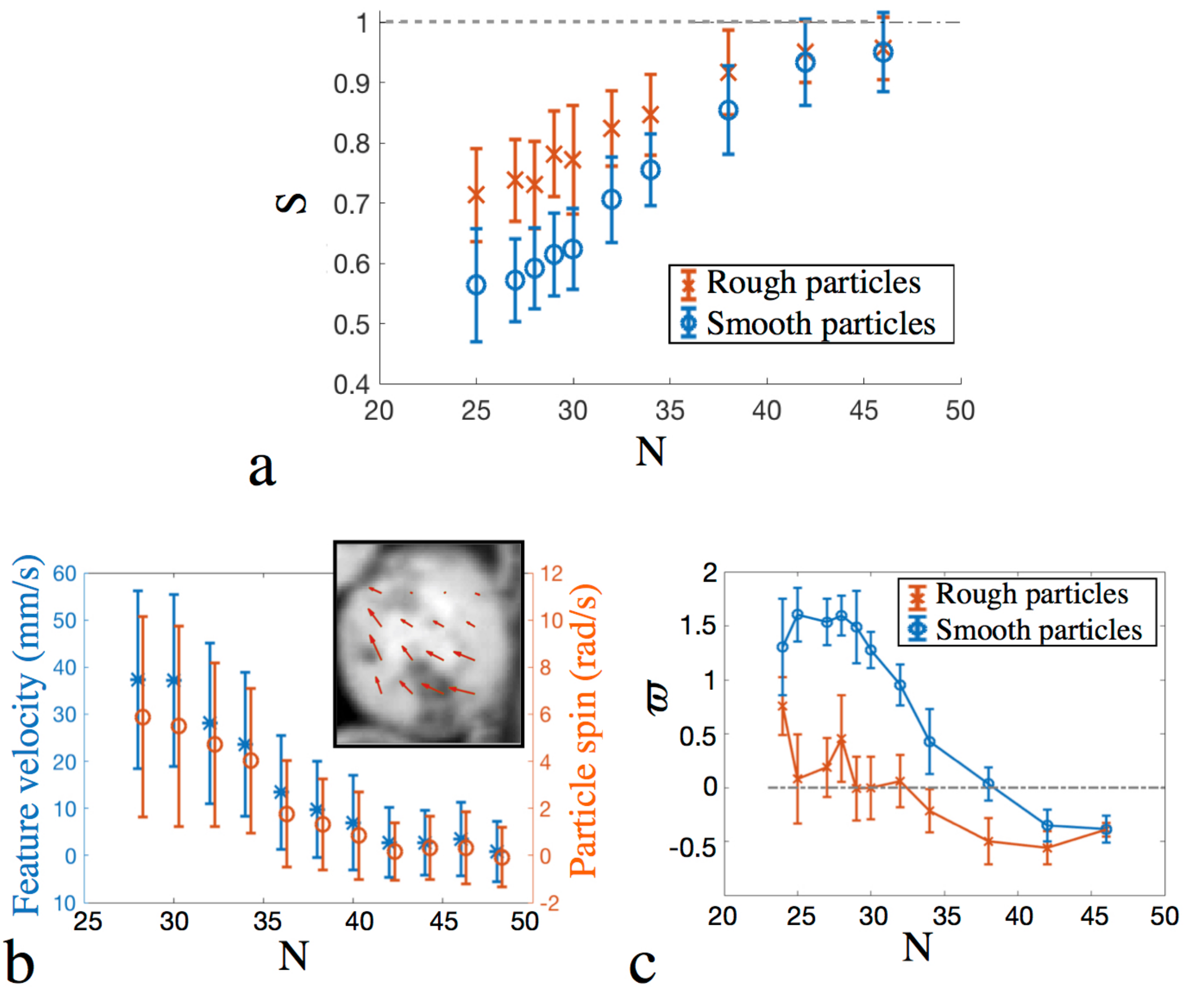}
\caption{
a: The slip parameter $s$ of the particles in the pinned region approaches 1 as $N$ is increased, corresponding to the pinned region particles moving more coherently with the moving wall. Additionally, the slip parameter $s$ of roughened particles along the wall of the pinned region is closer to 1 (non-slip conditions) than for smoother particles, meaning that roughened particles also move more coherently with the moving wall in the pinned region. b: The local rotation of individual particles along the wall in the pinned area decreases with increasing $N$, a result of increasing frustration between the particles. Inset shows snapshot of a particle with overlaid Particle Image Velocimetry (PIV) vectors. We use PIV on the surface features of the particles to determine general particle rotation about any axis (blue) and spinning about an axis going into the dish through the particle's center (orange). c: Particles with roughened surfaces, and therefore increased friction, transition to counterrotation at a lower $N$ than for smoother particles. \label{fig4}}
\end{figure}

At this point it is insightful to 
consider a  minimal model for our swirling system:  a single rigid ball of radius $R_\text{ball}$ swirled in a container of radius $R_\text{cont}$ (SI Video). 
%
%
%
A steady-state for the ball in the M-frame occurs when it is near the bottom-left of the container (near M), rotating at constant angular velocity $\varpi_b$ with the torque along the frictional boundary balancing the centrifugal force.
If we define a dimensionless parameter $\gamma=\frac{|\varpi_b|}{\omega}$, then by \eqref{labtoM}, $\gamma<1$ corresponds to rotation in the lab frame and $\gamma>1$ corresponds to counterrotation. 
The value of $\gamma$ depends on the slip conditions between the ball and the container wall, which interpolate between two limiting cases: perfect no-slip (strong friction) and perfect slip (no friction). 
When friction is strong, the ball rolls commensurately on the wall of the container without slipping, and  $\gamma=\gamma_{c}=\frac{R_\text{cont}}{R_\text{ball}}$. Since $R_\text{cont}>R_\text{ball}$, the ball thus counterrotates in the lab frame. 
Conversely, when there is no friction, the ball cannot rotate at all due to perfect slip boundary conditions and $\gamma=0<1$, corresponding to rotation. In between the two extremes, the degree of wall-slip can be measured by a slip parameter $s=\frac{\gamma}{\gamma_{c}}=\frac{\varpi_M}{\omega}\frac{R_\text{ball}}{R_\text{drum}}$, which varies from $s=0$ for perfect slip conditions, to $s=1$ for perfect no-slip conditions.

While the single-particle model captures much of why a swirled particle cluster can either rotate or counterrotate, the details are more subtle since an ensemble of particles rarely behaves exactly like a single rigid object. Even in the M-frame, where the dynamics are in steady state, the instantaneous angular velocity of any single particle $\varpi_i(t)$ depends on where it is in the dish, as shown in Fig \ref{fig3}a. It is therefore more appropriate to refer to a particle's instantaneous velocity as $\varpi_i(\theta)$. The change in $\varpi_i(\theta)$ with $\theta$ can largely be characterized by the loose and pinned regions.
In the loose region, the value of $\varpi_i(\theta)$ increases and decreases in a quantitatively similar manner for all values of $N$, with the only difference being the location on the dish at which this peak occurs. However, in the pinned region, the average value of |$\varpi_i(\theta)$| consistently increases with increasing $N$. 
Therefore, the pinned region is the major contributor to the different behaviors seen across varying $N$'s.
Comparison of $\varpi_i(\theta)$ in the pinned region to the velocity of the moving boundary reveals that the particles in the pinned region lag behind the moving boundary for low values of $N$, but approach the speed of the boundary at high $N$, as shown in Fig \ref{fig3}b.
Therefore, the average angular velocity of the cluster is dictated by the interaction of the particles in the pinned region with the container’s wall, in analogy to the slip-condition $s$ of the single-particle model. 

Indeed it is possible to define an analogous slip parameter for the swirling cluster,
$s=\frac{\gamma}{\gamma_{c}}=\frac{\varpi_M}{\omega}\frac{P_\text{ball}}{P_\text{cont}}$, where the radii of the rigid ball and container have been replaced by their respective perimeters, $P_\text{ball}$ and $ {P_\text{cont}}$, to account for the deformability of the cluster. 
Accordingly, 
the slip parameter of the pinned particles in our system increases from approximately 0.5 at $N=25$ to nearly 1 at the highest values of $N$, as shown in Fig \ref{fig4}a. 
Similarly, our experimentally swirled cluster behaves increasingly similarly to a model particle with no-slip boundary conditions as $N$ is increased, as shown in Fig \ref{fig1}b, confirming that counterrotation is due to increasingly no-slip boundary conditions.

The effective slip parameter $s$ is set by the dynamics of the individual particles along the wall within the pinned region,
but what determines the dynamics in the pinned region?
At low $N$, the particles at the wall are free to spin and roll locally, falling behind the moving wall and effectively reducing the particle cluster's friction with the container. 
This effective slip of the cluster on the wall corresponds to $s\ll1$, and the cluster rotates in the lab frame. 
As $N$ increases, the particles in the pinned region pack together more tightly, and friction between the particles dominates. When interparticle friction is strong between two contacting particles, they must spin about their own axes in opposite directions, dictating antiferromagnetic-like interactions for spinning. For six-fold packing, as we often see in the pinned region, this results in geometric frustration, prohibiting any of the particles from rolling or spinning freely at high $N$, as shown in Fig \ref{fig4}b. Similarly, particles in strong contact cannot advance as a tight single-file due to geometric frustration, and can only roll on the ground side-by-side. The inability to individually roll causes the  particles to effectively stick to the container wall, resulting in less effective wall-slip ($s\approx1$).
Therefore, the increased frictional effects at high $N$ cause the dense particle cluster to roll, or rather treadmill, on the container wall, resulting in counterrotation.

If friction is indeed what drives counterrotation, one would expect that increased frictional effects achieved by alternative means would also promote counterrotation. 
Indeed, sandpaper-roughened particles transition to counterrotation 
at $N=28$ as compared to $N=36$ for smooth particles, as shown in Fig \ref{fig4}c, confirming the importance of friction to counterrotation. Furthermore, the calculated slip parameter $s$ for the roughened particles is consistently closer to 1 than for the smooth particles, as shown in Fig \ref{fig4}a, affirming that counterrotation is associated with increasingly no-slip boundary conditions. 

Further experimentally testing the importance of friction to counterrotation is a challenge since it is difficult to systematically fine tune or completely eliminate the friction.
 We therefore turn to numerical simulations, which offer the unique advantage of adjusting physical constants that are impossible to change experimentally.

\begin{figure}[h]
\includegraphics[width=1\linewidth]{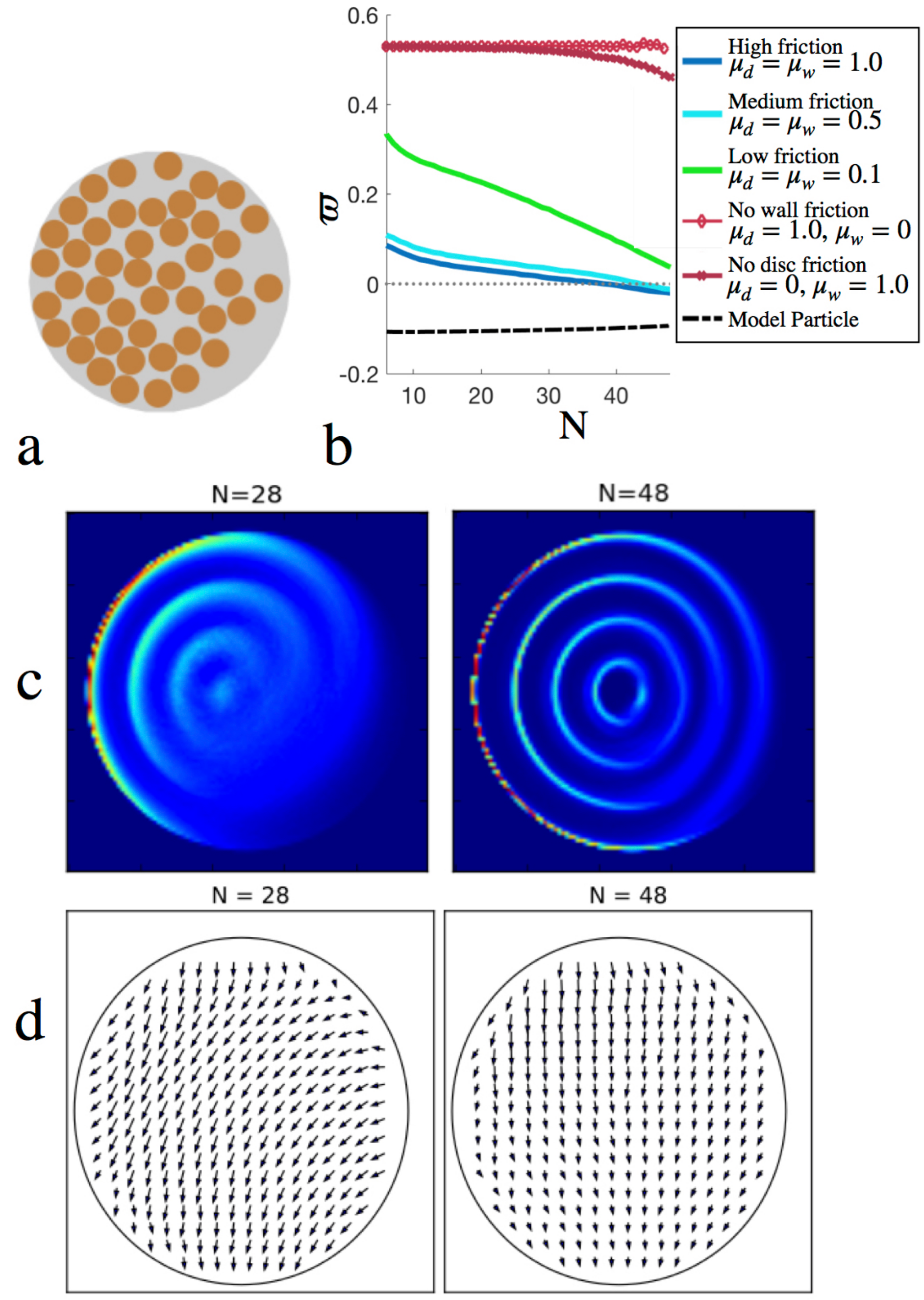}
\caption{
a: Snapshot of the simulated discs. b: The particle cluster transitions from rotation to counterrotation as $N$ is increased when friction is present. This rotation-counterrotation transition point occurs at higher $N$ when friction is decreased. When either disc-disc or disc-wall friction is completely eliminated, the system never transitions to counterrotation. Here "High friction" is $\mu_d=\mu_w=1.0$, "Medium friction" is $\mu_d=\mu_w=0.5$, and "Low friction" is $\mu_d=\mu_w=0.1$. When one friction is completely turned off, the other friction is set to 1.0.  c: Density histogram of the discs for a rotating ($N=28$) and counterrotating ($N=48$) case. d: Quiver plots showing the average local disc velocity deviation from the container for a rotating ($N=28$) and counterrotating ($N=48$) case.  \label{fig5}}
\end{figure}

We numerically simulate a system where $N$ two-dimensional discs with radius $r=1$ are swirled in a circular container with radius $R=8.6$, as shown in Fig \ref{fig5}a. The container is translated around a polygonal path with 30 sides and amplitude $A=0.96$, approximating a circle while allowing particle-wall collisions to be solved analytically. The behavior of the particles is simulated using an event-driven method, with the particles' linear and angular velocities updated every collision and otherwise determined via Newton's equations. Such a method allows us to exactly solve the dynamics up to floating-point precision, with the minimal number of parameters and ingredients in the model. Collisions are perfectly elastic in the normal direction, and in the tangential direction are subject to frictional impulses derived from Coulomb's law, using a coefficient of friction $\mu_d$ for particle-particle collisions and $\mu_w$ for particle-wall collisions \cite{r11} (see SI for Methods).

The average angular velocity $\varpi$ for the simulations is qualitatively similar to those measured in the experiments, exhibiting a rotation-counterrotation transition with increasing $N$, as shown in Fig \ref{fig5}b. 
The M-frame density histograms and relative angular velocities are also qualitatively similar to those of the experiment, as shown in Fig \ref{fig5}c and Fig \ref{fig5}d. 
The qualitative similarity between the experiments and simulations is striking, given that the simulations are significantly simplified -- notably, there is no friction with the substrate, no three-dimensional rolling effects, and no normal damping during collisions, so the discs never actually stick to each other. Therefore, these additional factors cannot be critical for observing the transition.

If the transition from rotation to counterrotation is indeed driven by friction both between the discs and with the container, then decreasing those frictions would require more discs to achieve the same effective fricton, and therefore the counterrotation transition should occur at a higher $N$. Indeed, decreased disc-disc and disc-wall friction cause the counterrotation transition to occur at higher $N$, as shown in Fig \ref{fig5}b. 
Furthermore,
completely turning off friction should eliminate the transition to counterrotation. We test these two cases separately.

We first test the importance of boundary friction by eliminating disc-wall friction ($\mu_w=0$), resulting in a frictionless boundary that serves only to contain the discs via hard-core-like elastic interactions. The discs may still frustrate each others' abilities to spin about their own axes, but without wall friction the discs are not encouraged to rotate commensurately with the wall.
In the lab frame, the cluster of discs behaves as a solid-like unit sloshing around the container without any internal individual spinning (see SI videos).
The angular velocity of the cluster therefore remains at that of the container regardless of $N$, as shown in Fig \ref{fig5}b.
In the M-frame, the discs form a rim at the boundary of the container with very little internal movement (SI video).
The steady-state of these particles arises from the centrifugal force pushing particles to the wall, much like sand settling under gravity.
However, without additional forcing from the wall, the angular velocity of the cluster cannot change, so the system never transitions to counterrotation, verifying that disc-wall friction is critical to induce counterrotation.

Next, we separately test the importance of disc-disc friction by eliminating it ($\mu_d=0$) and bringing back disc-wall friction. In this system, the outer discs are accelerated by frictional collisions with the wall and spin about their own axes quickly. However, with no disc-disc friction, none of this spinning is transferred to the inner discs except through particle exchanges, so individual discs spin with no coherence. The cluster of discs is loose and gas-like, unlike the rigid body appearance of the cluster when all friction is present (SI video).
With no disc-disc friction the average angular velocity remains close to the container velocity, except at very high $N$ where it drops slightly.
The system never transitions to counterrotation, verifying that disc-disc friction is also critical to induce counterrotation.

In conclusion, we perform experiments and numerical simulations of swirled granular media and identify the minimal ingredients for the particles and their interactions necessary to induce counterrotation.
We use a frame of reference in which the dynamics of swirled granular media are at steady state, the M-frame. In this frame of reference we introduce a minimal model, where the granular cluster is replaced by a single ball within a rotating drum. This system can capture the observed dynamics by only tuning the amount of slip $s$ that the ball experiences with the wall of the rotating drum.
When the ball experiences strong slip on the wall ($s\ll1$), the system rotates in the lab frame, while minimal slip ($s\approx 1$) corresponds to counterrotation in the lab frame.
In the granular system, $s\ll1$ is the result of individual particles falling behind the moving wall while rolling freely along the wall and the ground, acting as bearings for the particle cluster against the wall. This is generally true at low $N$ when the effective friction is minimal, and corresponds to rotation in the lab frame.
On the other hand, $s\approx 1$ occurs when the individual particles along the wall move with the wall due to high packing densities and geometric frustration. This is generally true for densely packed particles at high $N$.
Particle-particle friction prevents closely-packed particles from spinning and rolling freely, converting the particle ensemble into a solid-like cluster. At the same time, particle-wall friction causes that solid-like cluster of particles to stick to the wall, causing the overall cluster to roll commensurately along the wall and counterrotate, much like a pancake in a swirling pan.

Our investigation has considered the dynamics of discrete particles but it would be interesting to model the system with continuum equations, coupling internal spinning to a continuum notion of vorticity.
Such equations have modeled related systems \cite{scriven,lubensky,vitelli} and could perhaps give insight into the sensitive interplay between pressure, vorticity, friction, and external forcing that leads to counterrotation.
While our system is more complicated than the others that have been studied as it has no additional symmetries that lead to simplified equations, one could still study these equations numerically in the M-frame where the external forces are stationary.

In our swirling system, the large scale translation locally drives individual particles to roll on the ground and spin on their neighbors and the boundary. The individual particles then interact with each other via frictional collisions, causing the motion of the particles to eventually coalesce into system-size rotation or counterrotation.
In addition to the swirling container and rotating drum systems, the behavior associated with interacting, individually-driven particles can give rise to phase changes and bifurcations in other physical systems.
Several studies have shown that interactions between actively moving or rotating objects, both self-propelled and system-scale driven, can lead to collective angular momentum changes and rich phase behaviors \cite{lubensky,glotzer,lushi,vitelli,wioland2016,lushi2014,workamp,moisy}.
Specifically, in our system the interaction is particle friction and geometric frustration.
Particle-particle rolling frustration is likely responsible for the rolling-jamming transition, resulting in an apparent discontinuous jump in effective friction when sheared layers of spherical marbles exceed a critical thickness \cite{r14}. The transition to solid-like behavior in these systems is driven by the inability of contacting particles to co-spin or roll in-line, distinct from solid-like behavior that is achieved via jamming \cite{r15,r16} or rigidity percolation \cite{r17}.

Finally, our observation that the transition to counterrotation can be manipulated by roughening the particles inspires a speculative but interesting analogy with the transition to turbulence in pipe flow, where a roughness dependent transition has also been observed \cite{r12}. However, unlike pipe turbulence, it is not clear if the transition to counterrotation of swirling particles exhibits a true critical phenomenon.

\begin{acknowledgments}
We are grateful to Tadashi Tokieda for introducing us to this phenomenon. This work was supported by the NSF (DMR-1420570).  M. H.-C. and J. P. R. were supported by US Department of Energy, Office of Science, Office of Advanced Scientific Computing Research, Applied Mathematics Program under Award No. DE-SC0012296. M. H.-C and J. P. R. thank Leif Ristroph for procuring tabletop experimental materials. S. M. R. and M.H.-C acknowledge support from the Alfred P. Sloan Foundation.
\end{acknowledgments}

\bibliography{swirlingMarbles}

\pagebreak
\widetext
\begin{center}
\textbf{\large Supplementary Information for ``Geometric frustration induces the transition between rotation and counterrotation in swirled granular media''}
\end{center}
\setcounter{equation}{0}
\setcounter{figure}{0}
\setcounter{table}{0}
\setcounter{page}{1}
\makeatletter
\renewcommand{\theequation}{S\arabic{equation}}
\renewcommand{\thefigure}{S\arabic{figure}}



\section{Video}
Included:\\
Experimental lab frame: rotating, stall, counterrotating\\
Experimental M-frame: rotating, stall, counterrotating\\
Numerical lab frame all friction: rotating, counterrotating\\
Numerical M-frame all friction: rotating, counterrotating\\
Numerical lab frame no disc friction: rotating, counterrotating\\
Numerical M-frame no disc friction: rotating, counterrotating\\
Numerical lab frame no wall friction: rotating, counterrotating\\
Numerical M-frame no wall friction: rotating, counterrotating\\
Minimal model M-frame: no slip and slip\\
Minimal model lab frame: no slip and slip


\section{Calculation of location that particles become pinned and free (Fig 3)}

\subsection{Becoming pinned}
Particles experience multiple collisions while transitioning from the loose to the pinned region. The angular location on the dish at which particles become pinned is therefore determined by the average angle at which these collisions occur. A collision event is assigned to a particle $p$ at time $t$ if the particle experiences a sufficiently sharp turn in its trajectory, calculated using a threshold on the second derivative of its coordinates $x(t,p)$ and $y(t,p)$. 

\subsection{Becoming free}
The angular location on the dish at which particles become free is determined by the angular position at which the particle no longer travels parallel to the dish edge. To determine this, we calculate the best fit slope for the particle's coordinates between the preceding 7 frames and the following 7 frames. When this slope deviates from the perpendicular to the particle's radial vector on the dish (or the particle's azimuthal vector) by more than 10 degrees, we mark this particle as becoming ``free.'' Mathematically:

$$\theta_{\text{free}} = \sum_{t=7}^{\substack{\text{all} \\ \text{frames-7}}} 
\sum_{p=1}^{\substack{\text{all} \\ \text{particles}}} 
\theta(t,p) * \text{boolean}(|{\theta(t,p)} - \arctan (-\frac{x(t,p)-x_{\text{dish center}}}{y(t,p)-y_{\text{dish center}}}) |> 10), $$ where
$$\theta (t, p) = \frac{\sum_{i=t-7}^{t+7}  (x(i,p) - \bar{X})(y(i,p) - \bar{Y})}{\sum_{i=t-7}^{t+7} (x(i,p) - \bar{X})^2},$$
and $$ \bar{X} = \frac{\sum_{i=t-7}^{t+7}  x(i,p)}{15}, \bar{Y} = \frac{\sum_{i=t-7}^{t+7}  y(i,p)}{15}.$$

\section{Calculation of particle frustration (Fig 4b)} 
Particle Image Velocimetry (using MATLAB's PIVlab) is run on the particles within the pinned region. The PIV interrogation windows are set such that a total of 16 velocity vectors is calculated within each particle. The velocity values ($\vec{v}_{\text{PIV}}$) are then used to calculated the following:

\subsection{Average velocity}
Average velocity is the average of the velocity vector magnitudes:
 $${\text{Average velocity}}=\sum^{\text{all vectors}} |\vec{v}_\text{PIV}|.$$

\subsection{Average particle spin}
Average particle spin takes into account only the azimuthal velocity, calculated by averaging the dot product of each PIV vector with its azimuthal vector (the azimuthal vector of the coordinates at which the PIV arrows have been calculated, taking the middle of the particle as the origin):

$$ {\text{Average particle spin}}=\sum_i^{\text{all vectors}} (\vec{v}_{\text{PIV},i} \cdot \frac{\text{azimuthal vector at point i}}{|\text{azimuthal vector at point i}|}) .$$



 \section{Numerical simulation method}

 \subsection{Overview}
 
 Our simulations consider $N$ identical discs of radius $R_{disc}$ in a circular container of radius $R_{container}$. We treat the container as a disc with a different radius so the number of discs in the simulation is $N+1$. The discs are labelled $0,1,2,\ldots, N$, where 0 indexes the container. The discs have mass $m$ and moment of inertia $I$. The container is assumed to have infinite mass and moment of inertia. The state of the system at any point in time is given by the positions of the discs and container $\{x_i\}_{i=0}^N$, $x_i\in \R^2$, the velocities of the discs and container $\{v_i\}_{i=0}^N$, $v_i\in \R^2$, and the angular velocities of the discs $\{\omega_i\}_{i=1}^N$ (the container does not rotate.) 
 
The container is ``swirled,'' translated along a path which is topologically a loop, at frequency $\omega$. We translate at constant speed along a path which is a regular polygon with $M$ sides and amplitude  $A$ (distance from center of polygon to vertex.) This path approximates a circle but allows us to solve analytically for disc-container collisions. 

The dynamics of the discs are assumed to satisfy Newton's equations everywhere except at collisions, so the discs travel with constant linear and angular velocity. Notably, they do not experience friction with the floor. When discs collide with each other or the container,  their linear and angular velocities are updated according to a ``collision law,'' a rule for updating velocities which is consistent with certain physical principles. 
We choose our disc-disc collisions to be perfectly elastic in the normal direction, and in the tangential direction apply an impulse consistent with kinetic Coulomb friction (we ignore static friction.) Our disc-container collisions are similar but we consider the container to have infinite mass and moment of inertia, so its velocity doesn't change after a collision. More details on the collision law are given in section \ref{sec:collision} below. 

The dynamics are implemented using an event-driven method. We consider three kinds of events: 
\begin{enumerate}[1.]
\item Two discs collide;
\item A disc collides with the container;
\item The container reaches a vertex of its polygonal path. 
\end{enumerate}
The method proceeds by repeating the following steps. 
\begin{enumerate}[(i)]
\item Calculate the times of all future events;
\item Determine the time of the event that happens the soonest;
\item Determine the positions and velocities of the system at that time and then update them using the rule associated with the event. 
\end{enumerate}
When none of the events occurs we may solve analytically for the positions and velocities of all the discs. When an event occurs we must change the velocities using the rule associated with that event. 

 \subsection{Finding the next event} \label{sec:nextevent}
 
In this section we describe how the times of future events are calculated, considering each type of event in turn.
 
\begin{enumerate}[1.]
\item We calculate the collision time $t_{ij}$ for each pair of discs $(i,j)$. 
 For some disc $i$ with position $x_i$ at the current time, the position at any time increment $t$ from the current event (if there have been no collisions) is 
 \begin{equation}
p_i(t) = x_i + tv_i\,. 
 \end{equation}
A pair of discs $(i,j)$ in isolation will collide when the distance between their centers equals the sum of their radii:
\begin{equation}\label{eq:dist}
|p_i(t) - p_j(t)|^2 = (2R_{disc})^2\,.
\end{equation}
This is a quadratic equation in $t$ with positive leading coefficient, whose roots give the collision times. If both roots are negative, then the discs will never collide on their present trajectories (though they may have collided in the past). If the roots have different signs, then the current distance between centers is less than $2R_{disc}$, and so the discs are overlapping; note that this should never happen if the simulation has been implemented correctly. If the roots are both positive, then the lesser of them is the time that the discs will collide. If the roots are not real, then the discs will never collide. Therefore we can set $t_{ij}$ to be the lesser of the roots if it is real and positive, or $t_{ij}=\infty$ otherwise.

We make one small change to the above to account for numerical errors. 
It is sometimes the case that two discs that have just collided are  
predicted to collide next by the program (i.e., their solutions to \eqref{eq:dist} are real and positive.) In theory, the parabola in $t$ from \eqref{eq:dist} should only have one root at the origin, but numerical errors may cause the roots to both be real and positive. When this has happened, the predicted next time of collision has always been less than $10^{-14}$. To prevent these errors from being mistaken for actual collision predictions, we require that a predicted time of collision be greater than $10^{-13}$, otherwise we ignore that prediction. The minimal time between collisions is around $10^{-11}$ and $10^{-9}$ with and without friction, respectively.

\item For collisions between discs and the container, the collision times are calculated in a similar way as they are for pairs of discs except now we use the quadratic equation 
 \begin{equation}
 |p_i(t)-p_0(t)|^2 =|R_{container}-R_{disc}|^2
 \end{equation}
and choose the greater of the roots. As long as the discs always stay within the container, this quadratic will always have real roots of different signs (except in the case that the container and the disc have exactly the same linear velocity.)

\item Calculating the time of the next update to the container's velocity is trivial, since the container reaches a vertex of its polygonal path at fixed time increments. 
\end{enumerate}

 
Our current method calculates the collision times of all pairs of discs, which is an $O(n^2)$ calculation. The method could be accelerated by using cell lists, making it an $O(n)$ calculation \cite{frenkel}. Our method was fast enough for the numbers of discs under consideration but simulating large numbers of discs would require cell lists.

\subsection{Collision law}\label{sec:collision}

\begin{figure}
\begin{center}
\begin{tikzpicture}
\draw (0,-1.5) circle (1.5cm);
\draw (0,1.5) circle (1.5cm);
\draw (0,2.2) node[align=center] {disc $j$};
\draw (0,-2.2) node[align=center] {disc $i$};
\draw[->] (2.2,1.5) arc[radius=1.25cm, start angle=-5, end angle=35] node[anchor=west] {$\omega_j$};
\draw[->] (2,-2.5) arc[radius=1.25cm, start angle=325, end angle=365] node[anchor=west] {$\omega_i$};
\draw[->] (0,0) -- (0,1) node[anchor=south] {$\hat{n}$};
\draw[->] (0,0) -- (1,0) node[anchor=west] {$\hat{r}$};


\end{tikzpicture}
\end{center}
\caption{Sketch of two discs at the point of contact. Vectors $\hat n$, $\hat r$ are the normal and tangent directions as described in the text. }\label{fig:collision}
\end{figure}
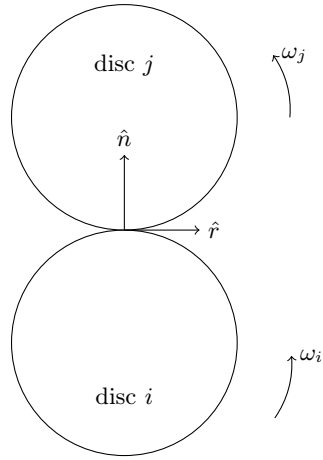

We now describe how to update the velocities of the discs after a collision. Our approach for relating tangential impulses to normal impulses is based on the calculations in \cite{WangMason1992}, although our equations are simpler since we consider only discs, and \cite{WangMason1992} considers arbitrary rigid bodies. These calculations can be justified asymptotically if the time over which the collision takes place is much shorter than a typical evolution time \cite{keller}. 

Suppose discs $i$ and $j$ have just collided so their boundaries are exactly tangent, as in Figure \ref{fig:collision}. Let $\hat n = (x_j-x_i)/|x_j-x_i|$ be the unit vector normal to the point of contact, pointing from $x_i$ to $x_j$, and let $\hat{r} = -(x_j-x_i)^\perp/|x_j-x_i|$ be the unit tangent vector at the contact. Here $(x,y)^\perp = (-y,x)$. For generality we let the discs have masses $m_i,m_j$ and moments of inertia $I_i,I_j$. 

Let $v_i,v_j,\omega_i,\omega_j$ be the velocities of the discs exactly when they contact but before the collision law has been implemented, and let $v_i',v_j',\omega_i',\omega_j'$ be the velocities after the collision law. 
Let $v_{i(j)}^{norm}$, $v_{i(j)}^{tang}$ be the normal and tangential components of each disc's initial linear velocity, i.e. $v_i^{norm} = v_i \cdot \hat n$, $v_i^{tang} = v_i\cdot \hat r$, and similarly for $v_j$. 
The velocity components are updated by increments $\Delta v_{i(j)}^{norm}$, $\Delta v_{i(j)}^{tang}$, $\Delta \omega_i$ giving updated velocities 
\[
v_i' =  (v_i^{norm}+ \Delta v_i^{norm})\hat{n}  + (v_i^{tang} + \Delta v_i^{tang})\hat{r} \,,\qquad
\omega_i' = \omega_i + \Delta \omega_i\,,
\]
and similarly for $v_j'$. 

The change in normal velocities is calculated assuming an elastic collision. 
Because we consider only discs, the normal force always points toward the centers of mass and does not mix with the tangential direction, so an elastic collision is equivalent to conserving energy and momentum conservation in the normal direction. (This is related to the equivalence of Newton's law of restitution, and the Poisson hypothesis, for discs, see \cite{WangMason1992}.)
The increments are
\begin{align*}
\Delta v_i^{norm} &= \frac{2m_j}{m_i+m_j}(v_j^{norm}-v_i^{norm})\\
\Delta v_j^{norm} &= \frac{2m_i}{m_i+m_j}(v_i^{norm} - v_j^{norm}) \;=\; -\frac{m_i}{m_j}\Delta v_i^{norm}\,.
\end{align*}

We wish to derive tangential increments from the Coulomb friction law for kinetic friction. Suppose the collision is not instantaneous, but rather takes place over a time interval of length $\tau$. Let $F_n(t)$, $F_f(t)$ be the normal and tangential components of the force acting on disc $j$ during the collision; $F_f(t)$ is the friction force. The forces acting on disc $i$ are equal and opposite, by Newton's third law.  The Coulomb friction law says that \cite{landau}
\begin{equation}\label{eq:coulomb}
F_f(t) = s_0\,\mu |F_n(t)| \,,
\end{equation}
where $\mu$ is the coefficient of kinetic friction.  
The above holds provided the relative velocity at the point of contact is nonzero; 
after it hits zero $F_f(t)=0$ and the relative velocity is zero for the remainder of the collision. 
The sign $s_0\in \{-1,1\}$ is chosen to ensure that friction reduces magnitude of the relative velocity of the points in contact. 
It is computed from the initial relative velocity $v_{rel}$ (the velocity of the contact point on disc $j$ minus the velocity of the contact point on disc $i$) as 
\begin{equation}
s_0 = -\sgn (v_{rel})\,, \qquad
v_{rel} =  v_j^{tang} - v_i^{tang}+ R_{disc}(\omega_i+\omega_j)\,.
\end{equation}

If we knew the normal force $F_n(t)$ and the length $\tau$ of the collision, then we could update the tangential velocities using Newton's second law by integrating \eqref{eq:coulomb}. We don't know either $F_n(t)$ or $\tau$ (and they probably depend sensitively on the details of the collision), but, crucially, we know the total \emph{impulse} in the normal direction, and this determines the total tangential impulse via \eqref{eq:coulomb}. 
Indeed, following \cite{WangMason1992}, we integrate \eqref{eq:coulomb} over the interval $[0,\tau]$ to obtain 
\begin{equation}\label{eq:impulse}
J_f = s_0\,\mu J_n\,,
\end{equation}
where $J_f = \int_0^\tau F_f(t)dt$, $J_n = \int_0^\tau F_n(t)dt$ are the impulses in the tangential and normal directions (for disc-disc collisions, by the choice of axes we have $F_n(t)\geq 0$.) 
Equation \eqref{eq:impulse} is valid provided the relative velocities never change sign; if they do the tangential impulse must be truncated as described below. 

The impulse in the normal direction $\hat n$ for disc $j$ is the change in the normal component of momentum,
\begin{equation}
J_n = m_j\Delta v_j^{norm}\,.
\end{equation}
The magnitude of the maximum impulse in the tangential direction is therefore 
\begin{equation}
J_f^{max} = \mu m_j \Delta v_j^{norm}\,.
\end{equation}
The actual tangential impulse is
\begin{equation}
J_f = s_0\, \min(J_f^{max}, J_f^{stick})\,,
\end{equation}
where $J_f^{stick}$ is the magnitude of the impulse required to make the relative tangential velocity between discs exactly zero (the superscript indicates the discs ``stick'' to each other in the tangential direction.) A formula for $J_f^{stick}$ will be derived in \eqref{eq:Jstick} below. 

Given $J_f$, the tangential increments are computed from the change in tangential and angular momenta:  
\[
m_j\Delta v_j^{tang}=J_f, \quad m_i\Delta v_i^{tang}=-J_f, \quad  
I_j\Delta \omega_j=R_{disc}J_f, \quad I_i\Delta \omega_i=R_{disc}J_f
\]
The tangential and angular velocities are updated as 
\begin{align}
 v_i^{',tang} = v_i^{tang} - \frac{J_f}{m_i} \,,&\qquad\qquad \omega_i' = \omega_i + \frac{R_{disc}J_f}{I_i}  \nonumber\\
 v_j^{',tang} = v_j^{tang} + \frac{J_f}{m_j} \,,&\qquad\qquad  \omega_j' = \omega_j + \frac{R_{disc}J_f}{I_j}
 \end{align}

The truncated impulse magnitude $J^{stick}_f$ is the impulse required to make the   relative tangential velocity between points in contact equal zero. It is found by solving equation
\[
v_j^{tang} + \frac{J_f^{stick}}{m_j} - \Big(v_i^{tang} - \frac{J_f^{stick}}{m_i}\Big) 
+ R_{disc}\Big(\omega_j + \frac{R_{disc}J_f^{stick}}{I_j} + \omega_i + \frac{R_{disc}J_f^{stick}}{I_i} \Big) = 0\,,
\]
resulting in 
\begin{equation}\label{eq:Jstick}
J_f^{stick} = \frac{\Big|v_j^{tang}-v_i^{tang} + R_{disc}(\omega_i+\omega_j)\Big|}{\frac{1}{m_i}+\frac{1}{m_j} + \frac{R_{disc}^2}{I_i} + \frac{R_{disc}^2}{I_j}}\,.
\end{equation}

 \bigskip
 
Processing collisions with the container is similar, except that we set $m_0=I_0=\infty$. In this case, the initial increment is
\[\Delta v_i^{norm} = 2(v_j^{norm} - v_i^{norm})\]
\[\Delta v_j^{norm} = 0\]
Since $\omega_0=0$ we do not need to worry about the change in sign in some of the formulas because the collision is interior.

\bigskip

We make one final remark, that one can modify this method to allow for non-elastic behavior in the normal direction. However, it is then possible for discs to stick to each other, after which one has to track the dynamics of the cluster. Collision laws between clusters can become arbitrarily complicated as one solves for the propagation of impulse between the discs in each cluster. Since the dynamics we seek to understand do not seem to require nonelastic collisions we did not consider adding this additional complexity.

\section{Simulation parameters}

The container's path is a polygon with $M=30$ sides, and the period of the swirl is 11.866 units of time. The container moves at one unit of distance per unit of time, so the radius of the polygon formed by the container's path is 1.892. Since the container's radius is 8.6, this makes the ratio of the radius of the container's path to the radius of the container equal to 0.22, which matches the experimental setup. 

The following are the default parameters used in the plots and videos. 
\begin{center}
    \begin{tabular}{ | l | l | p{5cm} |}
    \hline
    Number of discs & $N$& variable; typically 28-48 \\ \hline
    Mass of disc & $m$ & 1 \\ \hline
    Moment of inertia of disc & $I$ &  10 \\ \hline
    Radius of disc & $R_{disc}$ & 1 \\ \hline
    Radius of container & $R_{container}$  & 8.6 \\ \hline
    Sides in container's polygonal path & $M$ & 30  \\ \hline
    Period of swirl & $\omega^{-1}$ & 11.866071 \\ \hline
    Number of events calculated &  & $10^6$ \\ \hline
    Disc/Disc Friction Parameter & $\mu_d$ & 1  \\ \hline
    Disc/Container Friction Parameter & $\mu_w$ & 1  \\ \hline
    \end{tabular}
\end{center}

To remove the effect of a certain kind of friction from the simulation, we set that friction parameter to 0. A ``low" friction value means it is set to 0.1, a ``medium" friction value means it is set to 0.5, and a ``high" friction value means it is set to 1.0. Note that setting both parameters to 0 will result in a crash as the discs' energies will blow up. 

Figure 5b in the main text was generated by taking the average over all events, for each $N$ between $5$ and $48$, of the total angular velocity of the discs about their center of mass over $10^6$ iterations. 

For Figure 5c, the M-frame was divided evenly into a 100x100 grid of cells, and the amount of times that a disc's center was found inside a cell was measured for $N=28$ and $48$. Cells that are colored red were found to contain disc centers very often. 

For Figure 5d, the M-frame was divided evenly into a 20x20 grid of cells, and the average linear velocity of discs found in each of the cells was calculated for $N=28$ and $48$.

\end{document}